\documentclass[prb,twocolumn,showpacs,superscriptaddress,groupedaddress,amsmath,amssymb]{revtex4-1}

\usepackage{graphicx}
\usepackage{dcolumn}
\usepackage{bm}
\usepackage{hyperref}
\usepackage{color}
\usepackage{ulem} 

\begin{document}

\title{
Microscopic evidence for spin-spinless stripe order with reduced Ni moments within $ab$ plane for bilayer nickelate La$_3$Ni$_2$O$_7$ probed by $^{139}$La-NQR
}

\author{Mitsuharu Yashima$^1$, Nina Seto$^1$, Yujiro Oshita$^1$, Masataka Kakoi$^{1,2}$, Hiroya Sakurai$^3$, Yoshihiko Takano$^3$, and Hidekazu Mukuda$^1$}

\affiliation{$^1$Graduate School of Engineering Science, Osaka University, Toyonaka, Osaka 560-8531, Japan}
\affiliation{$^2$Department of Physics, Osaka University, Toyonaka, Osaka 560-0043, Japan}
\affiliation{$^3$National Institute for Materials Science (NIMS), Tsukuba, Ibaraki 305-0047, Japan}

\date{\today}

\begin{abstract}
The intrinsic electronic properties of La$_3$Ni$_2$O$_{7}$ have been selectively investigated by nuclear quadrupole resonance (NQR) at the La(2) site outside the NiO$_2$ bilayers. The La(2)$_{\rm a}$ site of the ideal La$_3$Ni$_2$O$_{7}$ is clearly distinguished from the La(2)$_{\rm b}$ site close to the local defects.
Below 150K, almost half of the intrinsic La(2)$_{\rm a}$ sites are dominated by a finite internal field within the $ab$ plane, while the other half are dominated by zero internal field. The result is fully consistent with the single spin-spinless stripe order of ($\cdots\uparrow\circ\downarrow\circ\uparrow\circ\cdots$), where the reduced Ni magnetic moments are parallel to the $ab$-plane.
Even for the La(2)$_{\rm b}$ site, the result is also explained within the same model by considering the inhomogeneous internal magnetic fields enhanced around the nearby defects such as oxygen vacancies. These findings provide unambiguous microscopic evidence for the single spin-spinless stripe order below 150 K  at ambient pressure.

\end{abstract}

\maketitle

\setcounter{section}{0}
\renewcommand{\thesection}{\arabic{section}} 
\section{\leftline{I\lowercase{ntroduction}}}
Remarkably high-$T_{\rm c}$ superconductivity (SC) with a maximum $T_{\rm c}$ of about $80$~K was reported for bilayer nickelate La$_3$Ni$_2$O$_{7+\delta}$ (La327) under high pressure in 2023.\cite{Sun_2023,Hou_replication_2023,Yanan-Zhang_replication,G-Wang_2024,Wang_La2PrNi2O7,Ko}
Since the formal valence of Ni is equivalent to Ni$^{2.5+}$$\,$($d^{7.5}$), the $3d_{x^{2}-y^{2}/3z^{2}-r^{2}}$ orbitals are partially occupied by 1.5 electrons. Thus, multiple degrees of freedom of spin/charge/orbital states in $d$ electrons are expected to play some essential roles for the emergence of SC phase under high pressure. In previous studies on La327($\delta \! \sim \! 0$) at ambient pressure, the anomalies in electronic states  below 100 $\sim$ 200 K has been reported, suggesting the existence of spin density wave (SDW) and/or charge density wave (CDW). \cite{Zhang, Wu, Liu2023, Liu2024, Hou_replication_2023, Yanan-Zhang_replication, G-Wang_2024, Chen_RIXS, Chen_muSR, Khasanov, Taniguchi, Dan_NMR, Kakoi, Meng, Ren, Gupta}
As possible candidates, the single spin-charge (spin-spinless) stripe order and the double spin stripe order have been proposed from the spectroscopies such as RIXS,\cite{Chen_RIXS} $\mu$SR,\cite{Chen_muSR,Khasanov} NMR,\cite{Dan_NMR} and RXS,\cite{Ren,Gupta} although the static magnetic order has not yet been reported by neutron scattering experiment.\cite{Neutron} 
A dual nature of electrons with spin and charge degrees of freedom may give rise to the complicated states associated with possible defects such as oxygen vacancies (O$_{\rm vac}$s) and stacking faults. It is highly desirable to determine the intrinsic magnetic and electronic properties of La327  by carefully distinguishing the extrinsic effects due to the defects. In the La327 crystal there are two crystallographically inequivalent La sites, one being La(1) between the NiO$_2$ planes and the other being La(2) outside the NiO$_2$ bilayers.
Nuclear quadrupole resonance (NQR) provides a unique opportunity to selectively probe the local electronic states at La(1) and La(2) sites, since each NQR frequency ($\nu_Q$) was deduced to be $\sim$ $5.6 \, (2.2)$ MHz for the La(2) ($\,$La(1)) site.\cite{Kakoi,Fukamachi_2001,Wang_La2PrNi2O7,Dan_NMR} In this Letter, we report the $^{139}$La(2)-NQR study on La327($\delta \sim$ 0), which provides unambiguous microscopic evidence for the single spin-spinless stripe order for the ideal crystal site below $T^* \sim$ 150K.

\section{\leftline{E\lowercase{xperimental}}}
Polycrystalline La327$\,$($\delta \! \sim \! 0$) was prepared by the solid state reaction method described elsewhere. \cite{Ueki}
X-ray diffraction ensures a single phase of the orthorhombic structure of La327$\,$($\delta \! \sim \! 0$) with lattice parameters $a$ = 5.3920\AA, $b$ = 5.4510\AA, and $c$ = 20.533\AA.\cite{Ueki} The oxygen deficiency $\delta$ is determined by thermogravimetric analysis.\cite{Ueki}
The $^{139}$La-NQR study was performed using a coarse powder of La327$\,$($\delta \! \sim \! 0$). For comparison, it was also carried out on the oxygen-deficient polycrystals La327$\,$($\delta \! \sim \! -0.03$) \cite{Adachi} and La$_4$Ni$_3$O$_{9.89}$$\,$(La4310)\cite{Nagata} synthesized by the different methods. 
The energy levels of the $^{139}$La nuclear spins ($I$ = 7/2) are split into four levels ($m \! = \! \pm1/2, \pm3/2, \pm5/2, \pm7/2$) by the nuclear quadrupole interaction, and thus three resonance peaks are observed for each La site in the NQR spectrum.
Here, the NQR frequency for the La($i$) is defined by $\nu_Q^{(i)} = 3eQV_{zz}^{(i)} \!/ [2I(2I-1)h)]$, where $V_{zz}^{(i)}$ is a principal value of the electric field gradient (EFG) tensor at the La($i$) site, and $Q$ is the electric quadrupole moment of $^{139}$La. 
The asymmetry parameter of the EFG $\eta^{(i)}$ is defined as $|V_{xx}^{(i)}-V_{yy}^{(i)}|/V_{zz}^{(i)}$. 
The $^{139}$La(2)-NQR spectrum was obtained at zero external field, focusing on the resonances at 2$\nu_Q$ ($\pm5/2 \leftrightarrow \pm3/2$) and 3$\nu_Q$ ($\pm7/2 \leftrightarrow \pm5/2$). The nuclear spin relaxation rate ($1/T_1$) was measured at 3$\nu_Q$ by the saturation recovery method \cite{Maclaughlin}.


\begin{figure}[htbp]
\hspace*{-0.7cm}
\includegraphics[width=9cm]{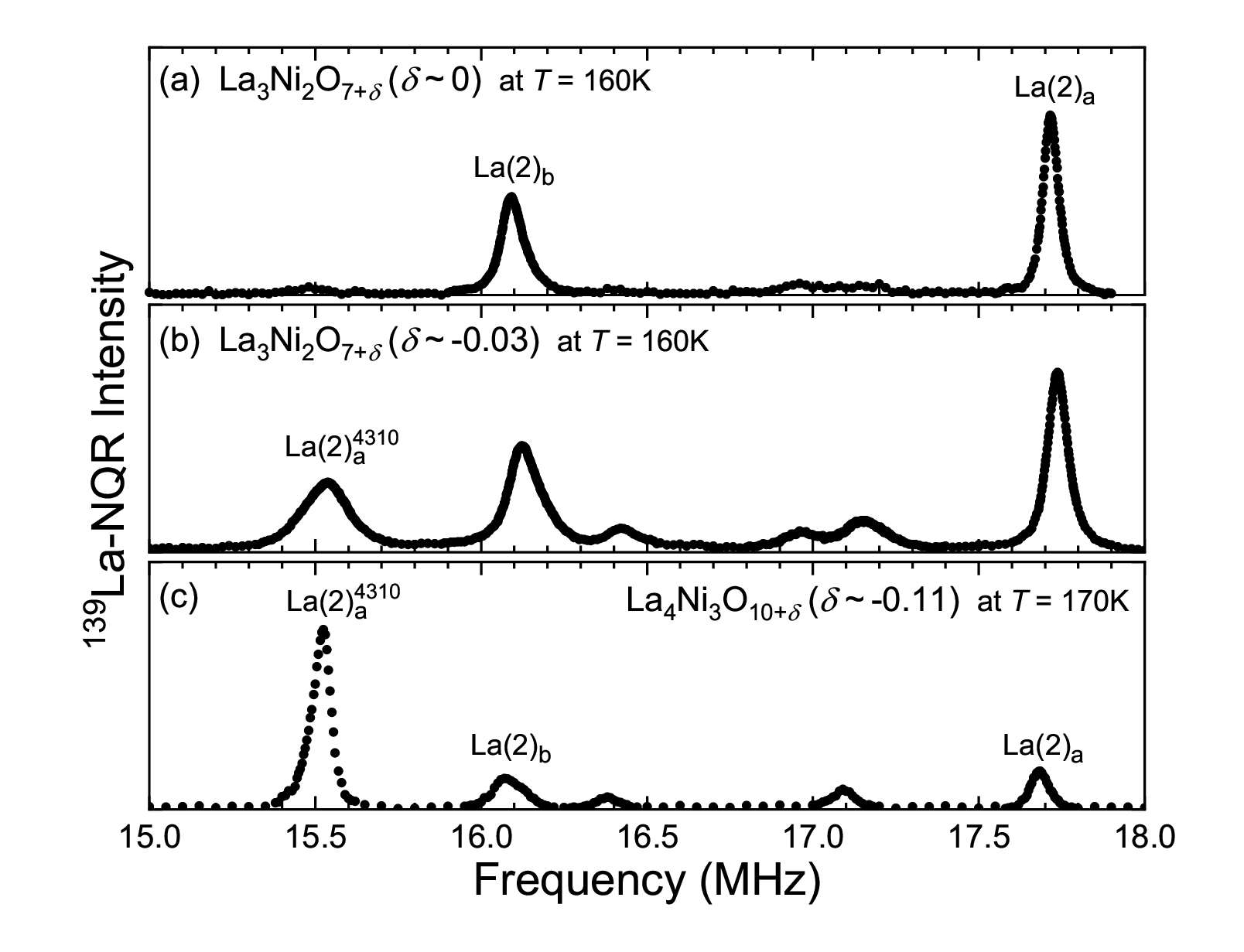}
\vspace*{-0.5cm}
\caption[]{\footnotesize (Color online) 
(a) NQR spectra at $3\nu_Q$ of La(2) above $T^*$ in (a) La327$\,$($\delta \! \sim \! 0$), compared to the results in (b) La327$\,$($\delta \! \sim \! -0.03$) and (c) La4310. In La327$\,$($\delta \! \sim \! 0$), two peaks, La(2)$_{\rm a}$ and La(2)$_{\rm b}$, are derived from the site of the ideal La327 and the one near O$_{\rm vac}$ at the inner apical site, respectively. The peak at $\sim \! 15.5$ MHz is the intrinsic La(2) site of La4310. The La327$\,$($\delta \! \sim \! 0$) sample in this study contains almost no stacking faults of La4310.}
\label{La2ab}
\end{figure}

\section{\leftline{R\lowercase{esults}}}
In a perfect crystal of La327 without any defects, we expect only one resonance peak from a single La(2) site in the NQR (3$\nu_Q$) spectrum above $T^*$ $\sim$ 150K in the paramagnetic state. However, as shown in Fig. \ref{La2ab}(a), even in La327$\,$($\delta \! \sim \! 0$), two distinct peaks are observed, labeled La(2)$_{\rm a}$ and La(2)$_{\rm b}$. The result is compared with the typical oxygen-deficient sample La327$\,$($\delta \! \sim \! -0.03$) in Fig. \ref{La2ab}(b) and La4310 in Fig. \ref{La2ab}(c). The peak of La(2)$_{\rm a}$ has the largest intensity with the narrowest linewidth in both La327 samples, and thus La(2)$_{\rm a}$ is assigned to the intrinsic La(2) site in the ideal La327 crystal without nearby defects. As for La4310, the largest peak with a narrow linewidth at $\sim15.5$ MHz is assigned to the La(2)$_{\rm a}^{4310}$ site of the ideal La4310 crystal without nearby defects. 
We emphasize that the La327$\,$($\delta \! \sim \! 0$) sample contains almost no stacking faults of La4310.

\begin{figure}[t]
\hspace*{-0.13cm}
\includegraphics[width=8cm]{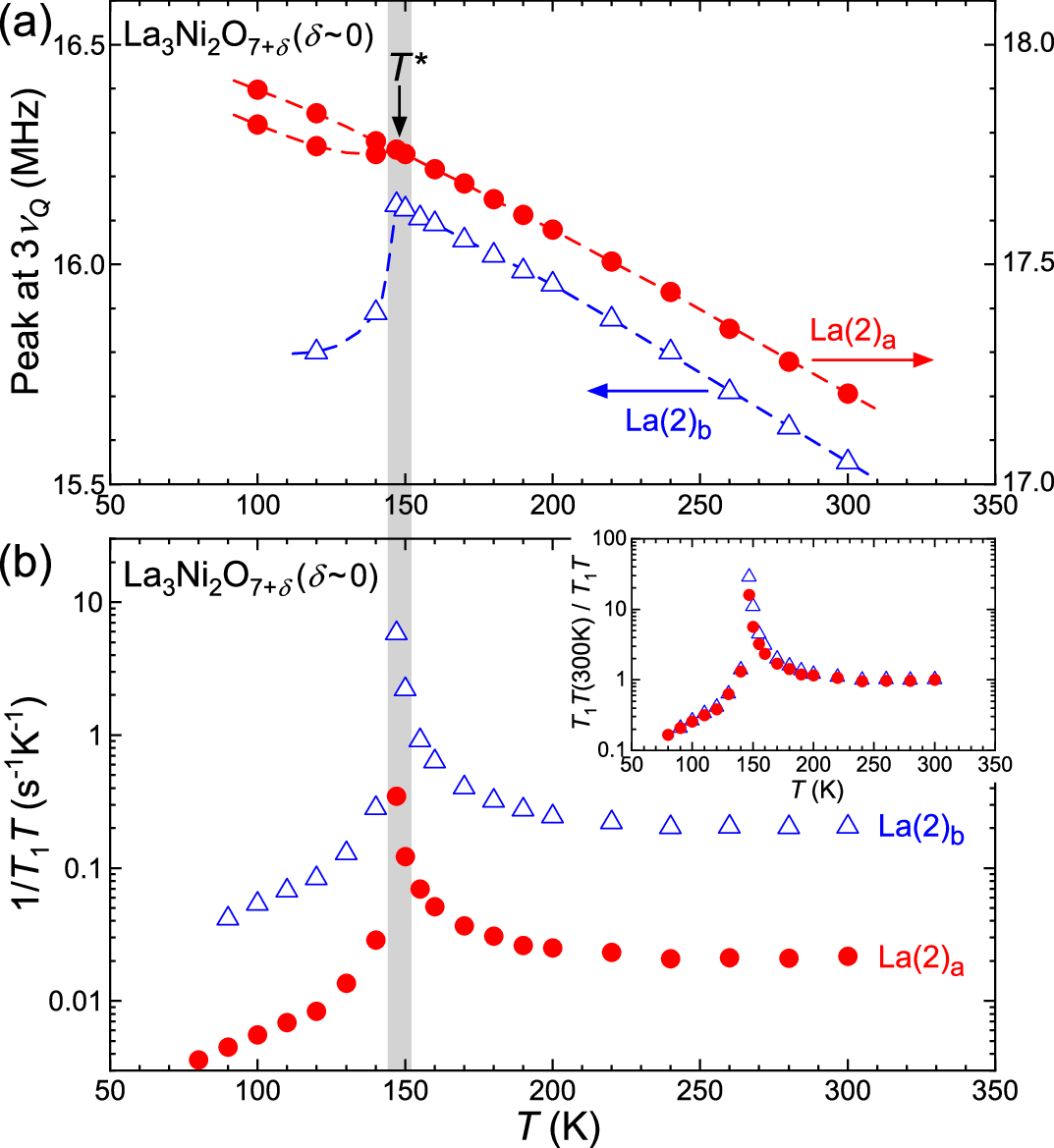}
\vspace*{-0.2cm}
\caption[]{\footnotesize (Color online) 
$T$-dependences of (a) peak frequencies at 3$\nu_Q$ and (b) $(T_1T)^{-1}$ for La(2)$_{\rm a}$ and La(2)$_{\rm b}$ of La327$\,$($\delta \! \sim \! 0$). The inset of (b) shows the identical $T$ dependences of $(T_1T)^{-1}$ normalized by the values at $T$ = 300 K for La(2)$_{\rm a}$ and La(2)$_{\rm b}$.
}
\label{T1}
\end{figure}

Next we consider the origin of the La(2)$_{\rm b}$ site in La327$\,$($\delta \! \sim \! 0$). From the analysis of the 2$\nu_Q$- and 3$\nu_Q$-spectra, we evaluate $\nu_Q^{\rm b}$ = 5.422 MHz and $\eta^{\rm b}$ = 0.155 for La(2)$_{\rm b}$ at $T$ = 160 K, which differ from the values $\nu_Q^{\rm a}$ = 5.952 MHz and $\eta^{\rm a}$ = 0 for La(2)$_{\rm a}$. As shown in  Fig. \ref{T1}(a), the $T$ dependence of the 3$\nu_Q$-peak frequency at La(2)$_{\rm b}$ is quite similar to that at La(2)$_{\rm a}$ above $T^*$. This indicates that they both belong to the same crystalline unit, which has almost the same $T$ dependence of the local EFG derived from the thermal shrinkage of the lattice. 
Furthermore, as shown in Fig. \ref{T1}(b), the $T$ dependence of $(T_1T)^{-1}$ at La(2)$_{\rm b}$ is exactly the same as that at La(2)$_{\rm a}$, although these absolute values are about 10 times different. In general, $(T_1T)_{\rm a,b}^{-1}$ is proportional to $|A_{\rm hf}^{\rm a,b}|^2 \chi''(q,\omega)$, where $A_{\rm hf}^{\rm a,b}$ is a hyperfine coupling constant at each La(2) site, and $\chi''(q,\omega)$ is a dynamical spin susceptibility.
The identical $T$ dependences of $(T_1T)_{\rm a,b}^{-1}$ are attributed to the $\chi''(q,\omega)$ coming from NiO$_2$ bilayers in common. Thus, the difference in the absolute values of $(T_1T)_{\rm a,b}^{-1}$ is mostly attributed to that in $A_{\rm hf}^{\rm a,b}$, which allows us to obtain the ratio $|A_{\rm hf}^{\rm b}/A_{\rm hf}^{\rm a}|\sim$ 3.06 from $[(T_1T)_{\rm b}^{-1}/(T_1T)_{\rm a}^{-1}]^{0.5}$. Considering these experimental facts, we assigned La(2)$_{\rm b}$ to the La(2) site near an O$_{\rm vac}$. It has been reported that most of O$_{\rm vac}$s exist at an inner apical site between NiO$_2$ planes.\cite{Ovacancy} The inner apical O$_{\rm vac}$ probably induces the anisotropy and inhomogeneity of EFG at the La(2)$_{\rm b}$ site, which has the finite $\eta^{\rm b}$ (= 0.155) and the broader linewidth than that of La(2)$_{\rm a}$ for each sample. 
In addition, the presence of O$_{\rm vac}$ gives a local distortion to its surrounding space and induces a change in the hybridization between Ni and La(2), probably leading to the modulation of $A_{\rm hf}$ at La(2) transferred from nearby Ni sites.

\begin{figure}[htbp]
\hspace*{-1.4cm}
\includegraphics[width=11.0cm]{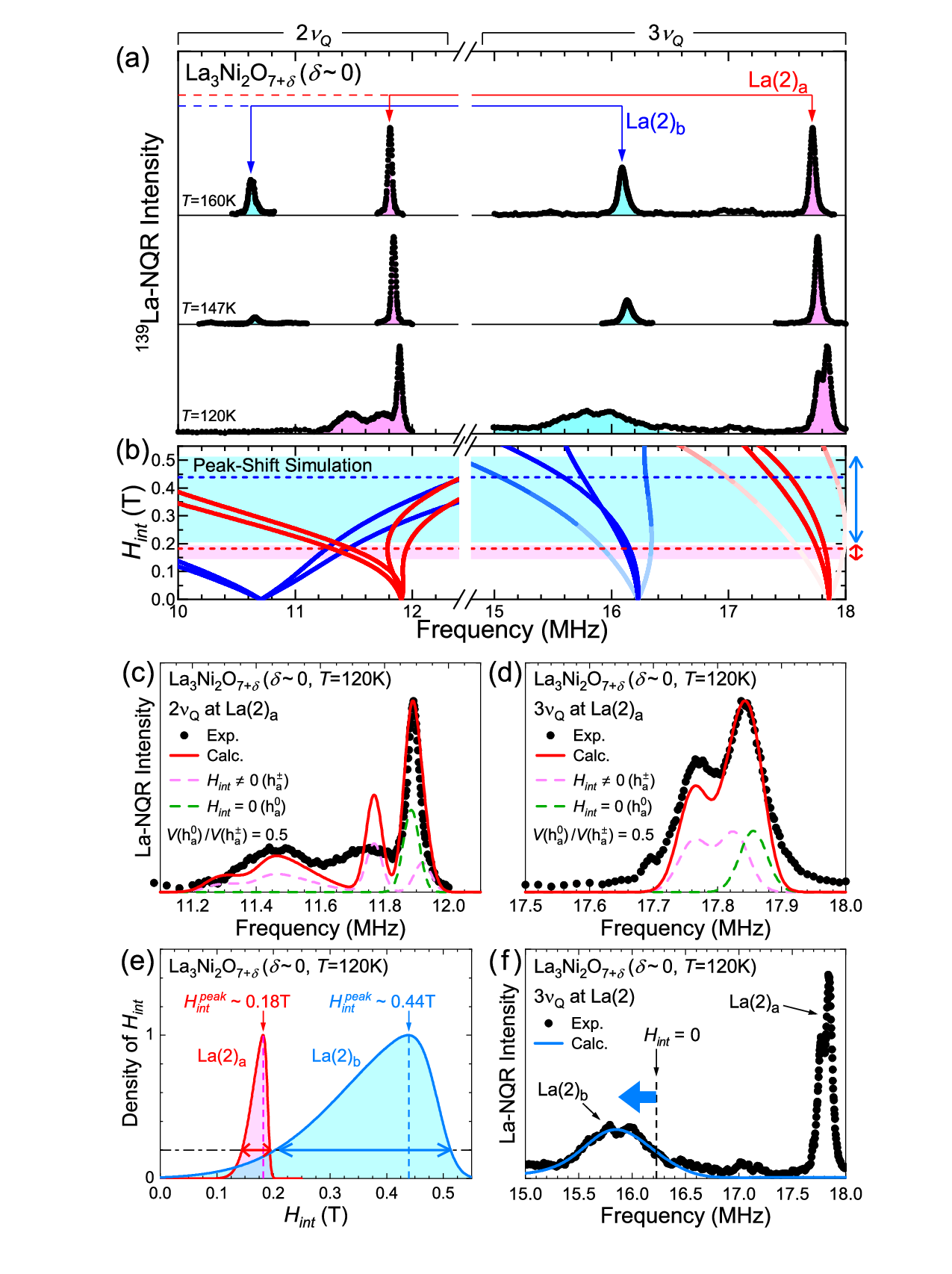}
\vspace*{-0.5cm}
\caption[]{\footnotesize (Color online) 
(a) La(2)-NQR spectra at 2$\nu_Q$ ($10 \! \sim \! 12$ MHz) and 3$\nu_Q$ ($15 \! \sim \! 18$ MHz) for La327$\,$($\delta \! \sim \! 0$) across $T^*$. 
(b) Simulation of peak frequencies as a function of $H_{\rm int}$ for La(2)$_{\rm a}$(red) and La(2)$_{\rm b}$(blue). 
(c) 2$\nu_Q$- and (d) 3$\nu_Q$-spectra for La(2)$_{\rm a}$ are roughly reproduced by the simulation (red line) superimposing two comparable components from the sites with finite $H_{\rm int}$ (pink dashed line) and zero $H_{\rm int}$ (green dashed line).
The intensities of two components are divided by two for clarity.
(e) Distributions of $H_{\rm int}$ used for the simulations in (c), (d), and (f), showing $H_{\rm int}^{\rm peak}$ and full width at 1/5 maximum for La(2)$_{\rm a}$ (red) and La(2)$_{\rm b}$ (blue).
(f) The 3$\nu_Q$-spectrum of La(2)$_{\rm b}$ is reproduced only by the site with finite $H_{\rm int}$ (blue), indicating no peak of La(2)$_{\rm b}$ with $H_{\rm int}$ = 0.
}
\label{Hint}
\end{figure}

Next we focus on the spin/charge ordered states below $T^*$ at La(2)$_{\rm a}$ for La327$\,$($\delta \! \sim \! 0$). Figure \ref{Hint}(a) shows the $T$ variations of the 2$\nu_Q$- and 3$\nu_Q$-spectra across $T^*$. As for La(2)$_{\rm a}$, the 2$\nu_Q$-spectrum around 11.9 MHz is largely split into three peaks below $T^*$, where two broad peaks shift to the lower frequencies, while the other one with a narrow linewidth remains around the frequency corresponding to zero internal field.
In contrast, the 3$\nu_Q$-spectrum at $\sim \! 17.8$ MHz shows a slight splitting into two narrow peaks. The $1/T_1$ values measured at all these peaks are almost the same, ruling out the possibility of phase separation. Furthermore, it cannot be attributed to the charge anomaly primarily: considering the relation $\nu_Q \propto V_{zz}$, the deformation of the 3$\nu_Q$-spectrum should be $\sim$1.5 times larger than that of the 2$\nu_Q$-spectrum if the charge anomaly such as CDW is a primary cause. This is inconsistent with our result. Therefore, to reproduce the spectra of 2$\nu_Q$ and 3$\nu_Q$ simultaneously, the peak-shift was simulated as a function of the internal magnetic field ($H_{\rm int}$) and the angle $\theta$ between $H_{\rm int}$ and $V_{zz}$. The red curves in Fig. \ref{Hint}(b) show the calculated peak-frequencies of 2$\nu_Q$ and 3$\nu_Q$ against $H_{\rm int}$, assuming the appropriate angle $\theta$ = 88.3$^\circ$. Here, the dark (light) tone on the colored curve represents the large (small) intensity expected in the spectrum. 
If $\theta$ were $\sim \,$0 or much smaller than 90$^\circ$, the 3$\nu_Q$ spectrum would be largely split into two peaks, one of which should be observed at a higher frequency side than the frequency expected for zero $H_{\rm int}$. 
However, in fact, no such higher frequency peak is detected, indicating that $\theta$ is close to 90$^\circ$.
The simulations are compared with the experiments for 2$\nu_Q$ and 3$\nu_Q$ in Figs. \ref{Hint}(c) and \ref{Hint}(d), respectively. As shown by the pink dashed curves, a small splitting and small shift to lower frequency observed for 3$\nu_Q$-peaks and a large splitting and large shift for 2$\nu_Q$-peaks are simultaneously reproduced by $H_{\rm int}$ with the largest weight ($\equiv H_{\rm int}^{\rm peak}) \sim$ 0.18 T and $\theta \sim$ 88.3$^\circ$. Here, a skew normal distribution of $H_{\rm int}$ shown in Fig. \ref{Hint}(e) was used, which is explained in detail in the Supplement(A). We have to consider not only the presence of above-mentioned La(2)$_{\rm a}$ sites with $H_{\rm int} \neq$ 0 ($\rm{h_ a^\pm}$) but also the presence of La(2)$_{\rm a}$ ones with $H_{\rm int}$ = 0 ($\rm{h_ a^0}$), where $H_{\rm int}$ is almost canceled out, as shown by the green dashed curves in Figs. \ref{Hint}(c) and \ref{Hint}(d). The volume fraction of La(2)$_{\rm a}$ with $H_{\rm int}$ = 0 [$V(\rm{h_ a^0})$] is roughly comparable to that with $H_{\rm int} \neq 0$ [$V(\rm{h_a^\pm})$]. The total calculated spectra shown by the red solid curves in Figs. \ref{Hint}(c) and \ref{Hint}(d) are superimposed by two comparable components for finite $H_{\rm int}$ (pink) and zero $H_{\rm int}$ (green) with $V(\rm{h_a^0})$$/$$V(\rm{h_a^\pm}) \sim$ 0.5. This result is consistently explained by the single spin-spinless stripe order model, as described later.

As for La(2)$_{\rm b}$, the 3$\nu_Q$ spectrum at $\sim \! 16.2$ MHz in Fig. \ref{Hint}(a) exhibits a large broadening and shift below $T^*$.
The blue curves in Fig. \ref{Hint}(b) are the peak-shift simulation at 2$\nu_Q$ and 3$\nu_Q$ for La(2)$_{\rm b}$ as a function of $H_{\rm int}$ under the same $\theta$ (= 88.3$^\circ$) as that for La(2)$_{\rm a}$. To explain the very broad spectrum from 15 to 16.5 MHz (see Fig. \ref{Hint}(f)), the $H_{\rm int}$ must be inhomogeneously enhanced at La(2)$_{\rm b}$. In fact, the spectrum in Fig. \ref{Hint}(f) is well reproduced by the simulation (blue), assuming $H{\rm _{int}^{peak}}\sim$ 0.44 T and its skew normal distribution shown in Fig. \ref{Hint}(e). 
It is noteworthy that the La(2)$_{\rm b}$ site with $H_{\rm int}$ = 0 was not observed at $\sim \! 16.2$ MHz, as seen in Fig. \ref{Hint}(f), indicating that there are few or no La(2)$_{\rm b}$ sites with $H_{\rm int}$ = 0. 
In this context, it is reasonable to explain that the 2$\nu_Q$-spectrum of La(2)$_{\rm b}$ at $\sim \! 10.6$ MHz in Fig. \ref{Hint}(a) is undetectably small, which is due to a huge broadening and wide splitting by the broad distribution of $H_{\rm int} = 0.2 \sim 0.5$ T, as expected by the simulation of Fig. \ref{Hint}(b).   
These features are quite different from those for La(2)$_{\rm a}$. 
Since $H_{\rm int}$ is generally given by a product of $A_{\rm hf}^{\rm a,b}$ and the Ni moment ($M_{\rm Ni}$), the difference in $H_{\rm int}$ for La(2)$_{\rm a,b}$ sites is attributed to the variation in $A_{\rm hf}$ and/or $M_{\rm Ni}$. The experimentally obtained ratio $H_{\rm int}^{\rm peak}(b)$/$H_{\rm int}^{\rm peak}(a) \sim 2.44$ is quite similar to $|A_{\rm hf}^{\rm b}/A_{\rm hf}^{\rm a}| \sim$ 3.06 obtained from the ratio of $(T_1T)^{-1}$. It is noteworthy that the magnitude of $M_{\rm Ni}$ on average does not change significantly around La(2)$_{\rm a}$ and La(2)$_{\rm b}$, although the local defect is present near La(2)$_{\rm b}$.

\begin{figure*}[tb]
\hspace*{-0.2cm}
\includegraphics[width=17.0cm]{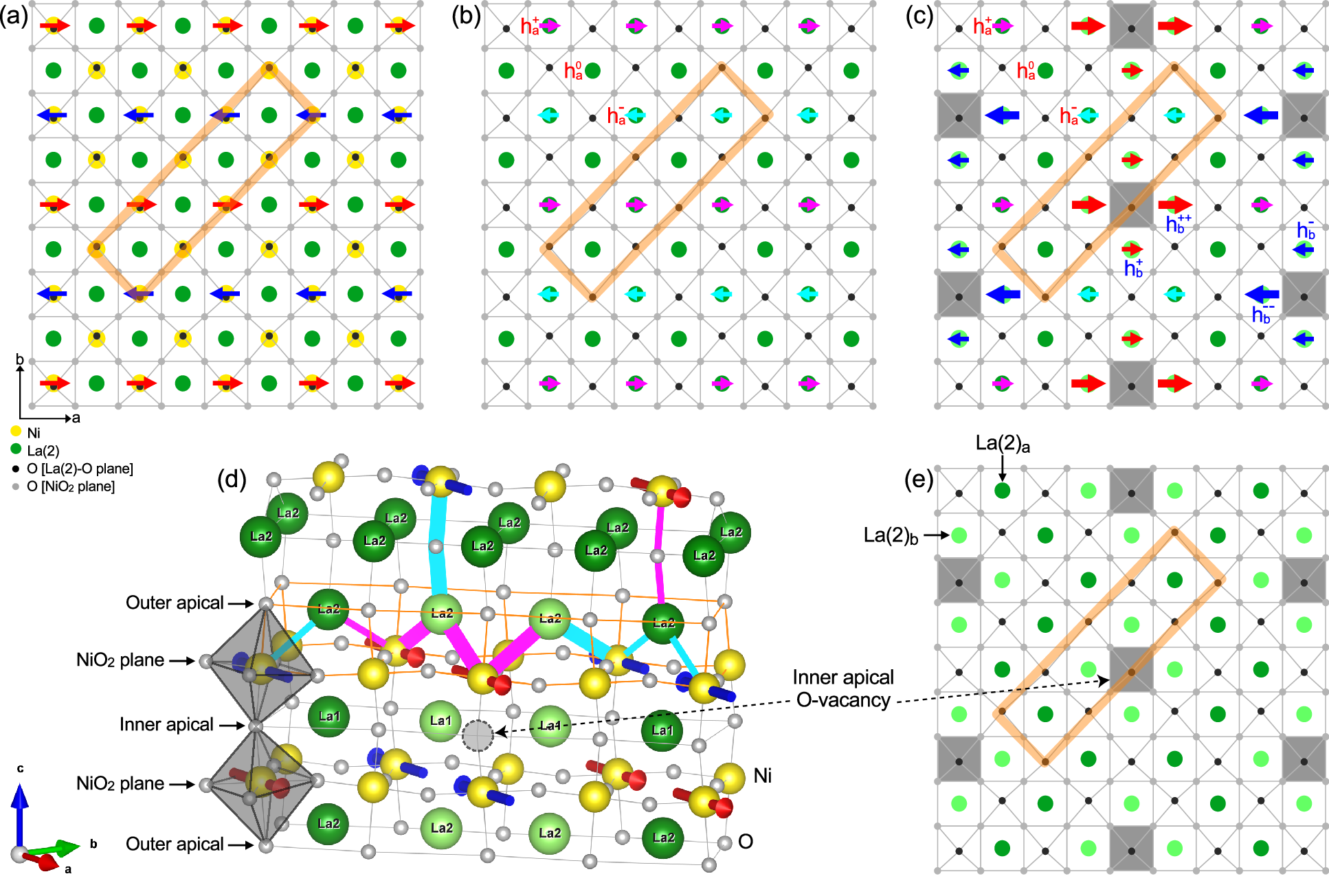}
\vspace*{-0.2cm}
\caption[]{\footnotesize (Color online) 
(a) Illustration of the single spin-spinless stripe order of $M_{\rm Ni}$s (red/blue arrows) antiferromagnetically aligned within the NiO$_2$ layers, consistent with this study. The La(2)-O plane is also projected onto this plane.  
(b) Spatial distribution of $H_{\rm int}$ at La(2)$_{\rm a}$ of the ideal La327, derived from the spin configuration of (a).    
(c) La(2)$_{\rm b}$ sites around O$_{\rm vac}$s are dominated by inhomogeneously enhanced $H_{\rm int}$ (see arrows of different sizes). 
(d) Structure of La327 with Ni moments (red/blue arrows) corresponding to the area enclosed by the orange rectangle in (a). 
The light green La(1) and La(2) sites indicate the nearest neighbor sites of O$_{\rm vac}$.
(e) Spatial distribution of La(2)$_{\rm a}$ (dark green) and La(2)$_{\rm b}$ (light green) around O$_{\rm vac}$s (gray) located on the Ni-spin channel above $T^*$, corresponding to (c) below $T^*$. The $H_{\rm int}$ at the La(2) is either parallel or antiparallel to the direction of $M_{\rm Ni}$ in this model, which is determined by the properties of $A_{\rm hf}$ composed of transferred and/or dipolar fields. 
For the sake of simplicity, we temporarily draw the case dominated by the dipolar fields.
}
\label{SDW}
\end{figure*}

\section{\leftline{D\lowercase{iscussions}}}
Figure \ref{SDW}(a) shows the illustration of the single spin-spinless stripe model, which is consistent with our NQR results. The $M_{\rm Ni}$s are antiferromagnetically aligned in the $ab$-plane as $(\cdots\uparrow\circ\downarrow\circ\uparrow\circ\cdots$) with the propagation vector $Q_{\rm SDW} = (0, 1/2)$, as shown by the red and blue arrows in Fig. \ref{SDW}(a) (see also Supplement(B) for the simplest case without any O$_{\rm vac}$s). This spin arrangement induces the comparable number of two La(2)$_{\rm a}$ sites with $H_{\rm int}=\pm$0.18 and 0 T in the $ab$ plane, corresponding to the $\rm{h_ a^\pm}$ and $\rm{h_ a^0}$ sites on the straight Ni-spin and spinless channels, respectively, along the $a$-axis in Fig. \ref{SDW}(b). 
We consider that the most possible direction of $M_{\rm Ni}$ is along the $a$-axis: The direction of $V_{zz}$ at La(2) is tilted by about 6$^\circ$ from $c$- to $b$-axis in association with the tilt of the NiO$_6$ octahedron, as shown in Fig. \ref{SDW}(d), and thus $\theta$ should naturally be close to $\sim \! 84^\circ$ if assuming $M_{\rm Ni} \parallel b$. However, the  experimentally obtained  $\theta$ ($\sim 88.3^\circ$)  is very close to 90$^\circ$, suggesting that the direction of $M_{\rm Ni}$ avoids the tilting direction of the octahedra ($b$-axis). 
We also exclude the cases of $M_{\rm Ni}\parallel c$ and non-parallel to $a$-axis within the $ab$ plane, because in such situations the $\rm{h_ a^0}$ site is eliminated.

Next we discuss the origin of the inhomogeneously enhanced $H_{\rm int}$ at the La(2)$_{\rm b}$ site.
As shown in Fig. \ref{La2ab}, the spectral intensity for La(2)$_{\rm b}$ is similar to that for La(2)$_{\rm a}$ in the paramagnetic state above $T^*$ for each La327 sample used in this study. If there is an O$_{\rm vac}$ at an inner apical site, as shown in Fig. \ref{SDW}(e), the neighboring four (eight for upper and lower sites) La(2)$_{\rm a}$ sites (dark green) are turned into La(2)$_{\rm b}$ sites (light green). 
Thus, the number of La(2)$_{\rm b}$ would be close to that of La(2)$_{\rm a}$,  assuming that the density of O$_{\rm vac}$ at the inner apical oxygen site is $\sim \! 1/8$ and that O$_{\rm vac}$s are separated from each other beyond the distance of the second nearest neighbor. 
Such a hypothetical La327$\,$($\delta \! = \! 1/8$) crystal with inner apical O$_{\rm vac}$s will induce the similar volume fraction for La(2)$_{\rm a}$ and La(2)$_{\rm b}$, which is consistent with this study. Considering this situation, as shown in Fig. \ref{SDW}(c), the nearest four (eight) La(2)$_{\rm b}$ sites around O$_{\rm vac}$s are further divided into two magnetically different sites below $T^*$, that is, the sites with strongly enhanced $H_{\rm int}$ ($\equiv \rm{h_ b^{\pm\pm}}$) and with moderate $H_{\rm int}$ ($\equiv \rm{h_ b^{\pm}}$), which appear on the Ni-spin and spinless channels around O$_{\rm vac}$s, respectively.
Even at the La(2)$_{\rm b}$ site on the spinless channel, denoted as $\rm{h_ b^{\pm}}$, the hyperfine field from a Ni spin next to an O$_{\rm vac}$ is plausibly different from that from a Ni spin without any O$_{\rm vac}$s, resulting in the imperfect cancellation of $H_{\rm int}$ from the surrounding Ni spins. 
It gives a reasonable explanation for the experimental facts that the value of $H_{\rm int}$ at La(2)$_{\rm b}$ is inhomogeneously enhanced, and no La(2)$_{\rm b}$ site with $H_{\rm int}$ = 0, which are accounted for within the same model used for the La(2)$_{\rm a}$ site.

We address the feature of $M_{\rm Ni}$ for the intrinsic La(2)$_{\rm a}$ site.  
The $H_{\rm int}$ was evaluated to be $\sim \,$0.18 T, which is one order smaller than $\sim \,$2 T at the La site of La$_2$NiO$_4$ (La214) in the antiferromagnetic (AFM) order by large $M_{\rm Ni}^{\rm 214} \sim 1.7 \, {\rm \mu_B}$ corresponding to a high spin state of Ni$^{+2}$$\,$($d^8$).\cite{Rodriguez-Carvajal,Wada} Since $A_{\rm hf}$ is different from that of La214, it is difficult to estimate the exact $M_{\rm Ni}$ of La327 so far.
However, at least the observation of the very small $H_{\rm int}$ in La327 suggests the significant reduction of $M_{\rm Ni}$ due to itinerant $d$-electrons in metallic La327, in contrast to the localized regime of insulating La214. According to the previous $\mu$SR experiment,\cite{Chen_muSR} $M_{\rm Ni} \sim 0.42 \, {\rm \mu_B}$ is anticipated if $M_{\rm Ni}$ is assumed to be within the $ab$ plane. 
It is further supported by our experimental fact that $M_{\rm Ni}$ is along the $a$-axis.
Such a reduction of $M_{\rm Ni}$ suggests that most of spins in the $d_{3z^2-r^2}$-orbital may be canceled by forming a spin-singlet like state, since the bonding-orbital part of the $d_{3z^2-r^2}$-band derived from the NiO$_2$ bilayers would be almost filled in La327.\cite{Sakakibara_PRL,Luo_theoryPRL,Yang_theoryPRB}Therefore, it is likely that the main amplitude of $M_{\rm Ni}$ originates from spins of the $d_{x^2-y^2}$-band and the mixed band of $d_{3z^2-r^2/x^2-y^2}$. If assuming that most of $M_{\rm Ni}$s would be derived only from spins in the nearly 1/4-filled $d_{x^2-y^2}$-orbital, the maximum of $M_{\rm Ni}$ is expected to be about $0.5 \, {\rm \mu_B}$, which is close to the $M_{\rm Ni}$ discussed above. 

Finally, we note that the CDW order at the intrinsic La(2)$_{\rm a}$ site cannot be clearly detected from the current study in the measured $T$ range from 100 to 300K. If the CDW order were present, it might be considerably weaker than the SDW order in this $T$ range. Therefore we use the term "spin-spinless" rather than "spin-charge" to correctly express the current result in this paper.
Our conclusion is different from the double spin stripe and double charge stripe order proposed microscopically by La(1)-NMR\cite{Dan_NMR} and La(2)-NQR\cite{Luo}, respectively.
In order to deduce the magnetic structure in La327,  two conditions must be satisfied, namely the existence of the $\rm{h_ a^0}$ site and the direction of $H_{\rm int}$ parallel to the $a$-axis. 
As shown in Fig. \ref{SDW}(d), the La(2) site is surrounded by four first-nearest-neighbor Ni sites   obliquely below La(2) and one second-nearest-neighbor Ni site above La(2). 
If the second-nearest-neighbor Ni site is not spinless, the cancellation of $H_{\rm int}$  at the La(2) becomes extremely difficult.  
Thus, the presence of the Ni-spinless channel is probably indispensable to explain the appearance of the $\rm{h_ a^0}$ site in the SDW order. 
Furthermore, a local symmetric spin configuration of four first-nearest-neighbor Ni sites  around La(2) is required for the good cancellation of $H_{\rm int}$. 
However, in the cases of the double spin stripe and double spin-charge stripe models, it is very difficult to obtain the innegligible amount of $\rm{h_a^0}$ sites due to the low local symmetry of the spin configuration around the La(2) site.
As for the direction of $H_{\rm int}$ in these double stripe systems, it should be fairly tilted from the $ab$-plane to the $c$-axis in many conditions, which is inconsistent with our NQR result.
Regarding to the single spin-spinless stripe structure, our NQR result is almost consistent with the RIXS,\cite{Chen_RIXS} $\mu$SR,\cite{Chen_muSR} and RXS\cite{Ren} studies. The large reduction of $M_{\rm Ni}$ in La327 deduced from the present NQR study is reasonably consistent with the undetectably small moment suggested by the neutron scattering experiment\cite{Neutron}. 
In addition, our NQR result is in good agreement with the spatial arrangement in which inner apical O$_{\rm vac}$s are located on the Ni-spin channel, not on the Ni-spinless channel (see the detail in Supplement(C)), indicating that the location of inner apical O$_{\rm vac}$s may not be completely random. If inner apical O$_{\rm vac}$s form such a pattern as the stripe, it is of interest in terms of how the pattern of O$_{\rm vac}$s affects the SDW order at low pressure and the superconductivity at high pressure.

\section{\leftline{C\lowercase{onclusions}}}
In summary, the $^{139}$La(2)-NQR measurement in La327 revealed the appearance of two intrinsic La(2)$_{\rm a}$ sites with zero and finite $H_{\rm int}$s below $T^* \sim$ 150K, which is consistently explained by the single spin-spinless stripe order model with the moderately reduced $M_{\rm Ni}^{\rm 327} \parallel a$ that is antiferromagnetically aligned by $Q_{\rm SDW}=(0,1/2)$.
Even for the La(2)$_{\rm b}$ site close to inner apical O$_{\rm vac}$s, these NQR results are mostly explained within the same model by considering inhomogeneous $H_{\rm int}$ enhanced around O$_{\rm vac}$s.
These results provide further insight into understanding the relationship with the high-$T_c$ states at high pressure.

\section*{\leftline{A\lowercase{cknowledgments}}}
This work was partially supported by the Takahashi Industrial and Economic Research Foundation and JSPS KAKENHI Grant No. JP24K01333. One of the authors (M. K.) was supported by JST SPRING, Grant No. JPMJSP2138 and by Kato Foundation for Promotion of Science, Grant No. KS-3614. Two of the authors (H. S. and Y. T.) were supported by World Premier International Research Center Initiative (WPI), MEXT, Japan, Grant No. JPMJSP2138.



\newpage %

\setcounter{figure}{0}
\renewcommand{\thefigure}{S\arabic{figure}} 

\centerline{{\bf Supplement}}

\subsection{Skew normal distribution used for the simulation of the asymmetric spectra }

In this simulation, the spectral components with $H_{\rm int}$ at 2$\nu_Q$ and 3$\nu_Q$ are calculated using the skew (asymmetric) normal distribution shown in Fig. \ref{Hint}(e). The skew normal distribution $S(x, \xi, \sigma, \alpha)$ is expressed with a complementary error function ${\rm erfc}(x)$ by
\[
S(x, \xi, \sigma, \alpha) = \exp \! \Big[- \Big( \frac{x-\xi}{\sigma} \Big)^2 \Big] \! \cdot {\rm erfc} \Big( \! -\alpha \frac{x-\xi}{\sigma} \Big),
\]
where $\xi$, $\sigma$ and $\alpha$ are a mean, a standard deviation of the distribution, and a shape parameter of the asymmetry of the distribution, respectively. This function with $\alpha$ = 0 is the same as the Gaussian (normal distribution) function. As for La(2)$_{\rm a}$, the calculated 2$\nu_Q$ and 3$\nu_Q$ spectra in Figs. \ref{Hint}(c) and \ref{Hint}(d) are simultaneously reproduced with the same parameters $(\xi, \sigma, \alpha)$ = (0.19$\,$T, 0.035$\,$T, $-6$), respectively. The distribution of $H_{\rm int}$ is shown by the red curve in Fig. \ref{Hint}(e). The peak position of this asymmetric $H_{\rm int}$ distribution ($H_{\rm int}^{\rm peak}$) is $\sim \!0.18$$\,$T. As for the La(2)$_{\rm b}$ site, the calculated spectrum (blue solid line in Fig. \ref{Hint}(f)) is reproduced by applying the parameters of the $H_{\rm int}$ distribution $(\xi, \sigma, \alpha)$ = (0.49$\,$T, 0.22$\,$T, $-6$). This gives $H{\rm _{int}^{peak}(b)} \sim$ 0.44$\,$T. The distribution of $H_{\rm int}$ is shown by the blue curve in Fig. \ref{Hint}(e). Both the amplitude and the distribution width of $H_{\rm int}$ for La(2)$_{\rm b}$ are much larger than those for La(2)$_{\rm a}$.

\newpage
\subsection{Single spin-spinless stripe order model : The simplest case for perfect crystal without any O$_{\rm vac}$s}
\begin{figure}[htbp]
\hspace*{-0.2cm}
\includegraphics[width=8.5cm]{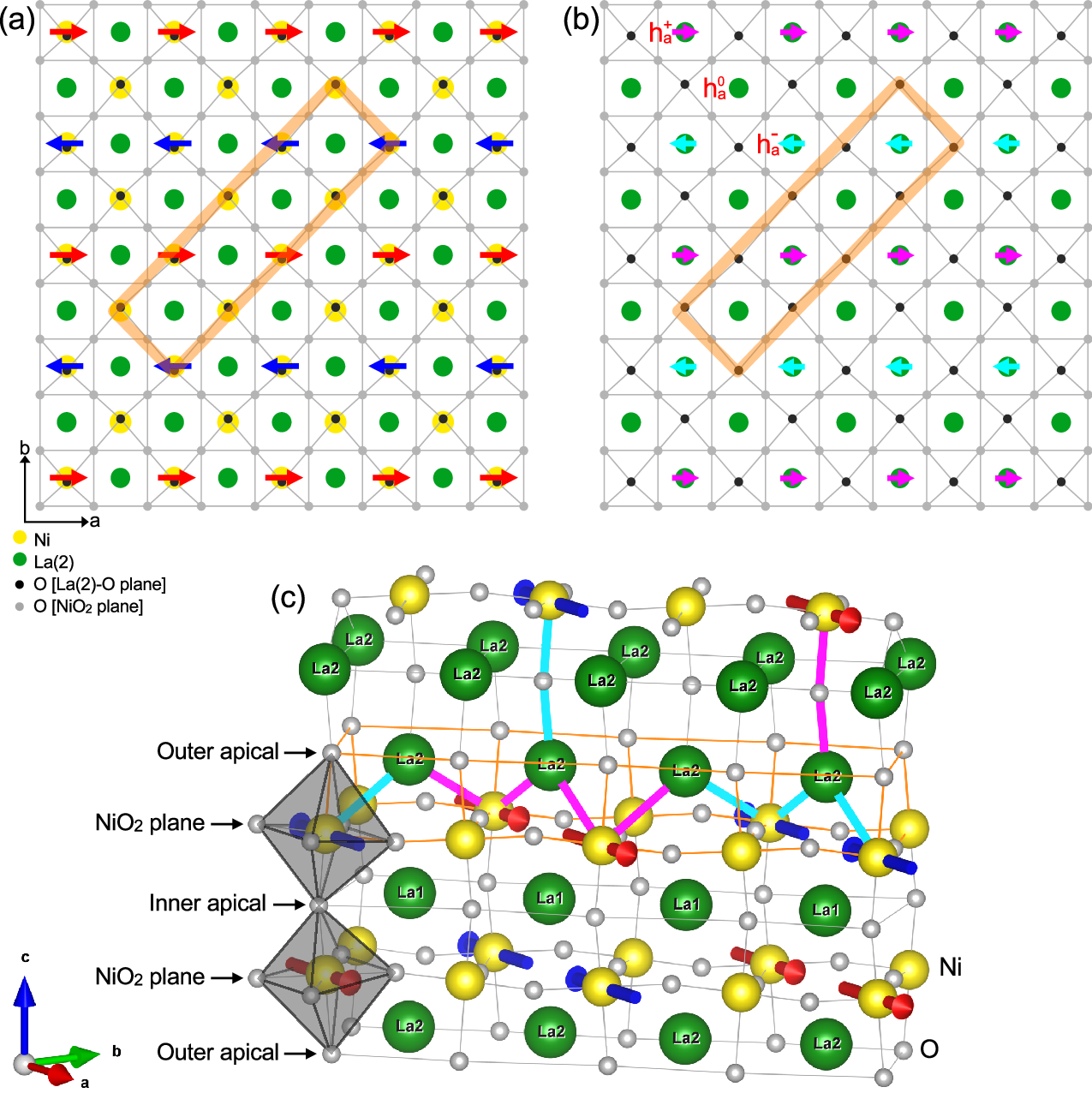}
\caption[]{\footnotesize (Color online) 
Single spin-spinless stripe order model for the simplest case in the perfect crystal without any O$_{\rm vac}$s. 
(a) Illustration of the single spin-spinless stripe configuration of $M_{\rm Ni}$s (red and blue arrows) antiferromagnetically aligned within the NiO$_2$ layers, consistent with this study. The La(2)-O plane is also projected onto this plane. 
(b) Spatial distribution of $H_{\rm int}$ at the La(2)$_{\rm a}$ site of the ideal La327, derived from the spin configuration shown in (a). Almost half of the intrinsic La(2)$_{\rm a}$ sites are dominated by a finite $H_{\rm int}$ within the $ab$ plane, while the other half are dominated by zero $H_{\rm int}$.
(c) Crystal structure of La327 including no O$_{\rm vac}$s with $M_{\rm Ni}$s (red and blue arrows).
As for the direction of $H_{\rm int}$ at the La(2) site, for the sake of simplicity, we temporarily draw the case where the dipolar fields are predominant.
}
\label{Perfect}
\end{figure}

\begin{figure*}[tb]
\hspace*{-0.2cm}
\includegraphics[width=17.0cm]{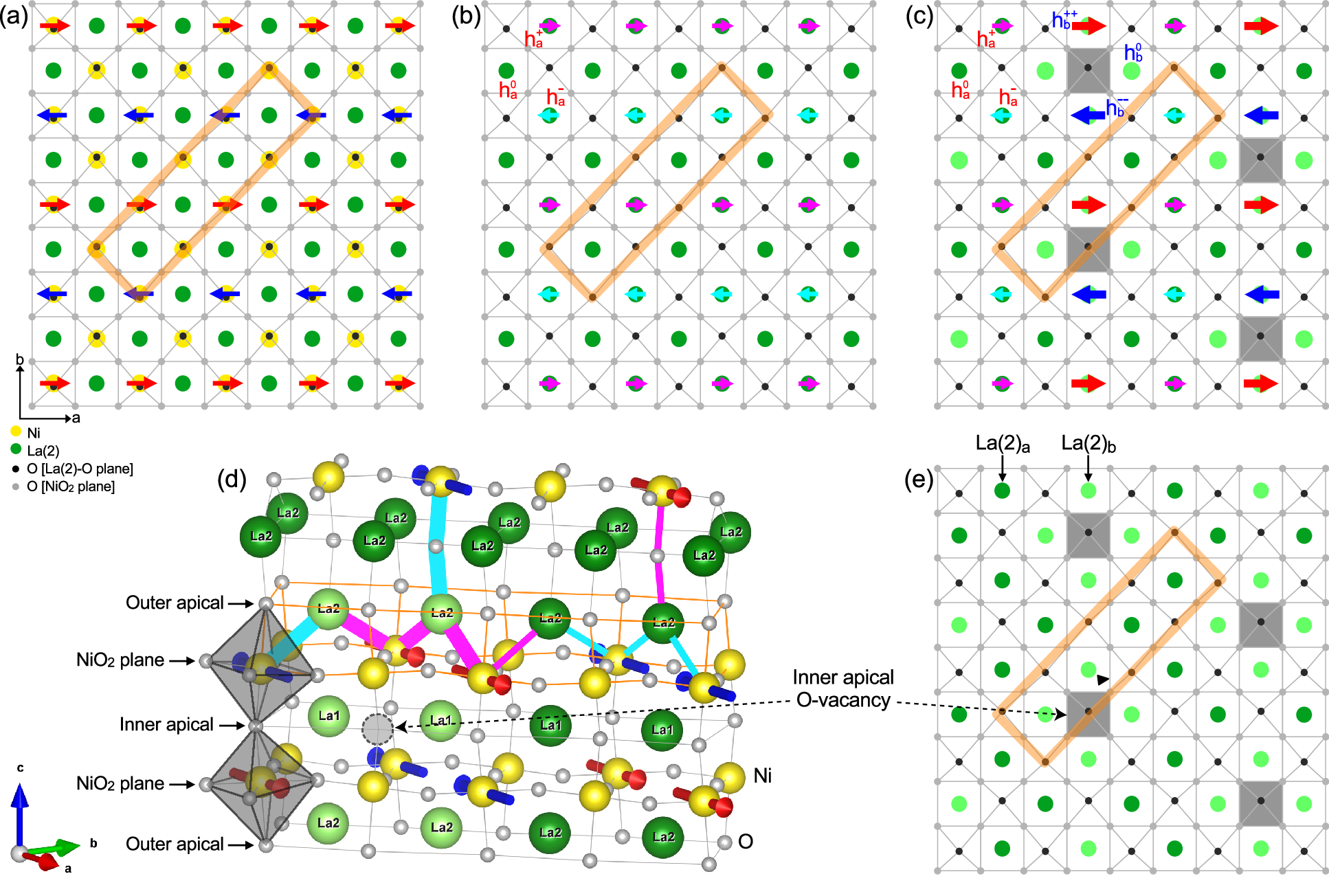}
\vspace*{-0.2cm}
\caption[]{\footnotesize (Color online)
(a) Illustration of the single spin-spinless stripe order of $M_{\rm Ni}$s (red/blue arrows) antiferromagnetically aligned within the NiO$_2$ layers, consistent with this study. The La(2)-O plane is also projected onto this plane. 
(b) Spatial distribution of $H_{\rm int}$ at La(2)$_{\rm a}$ of the ideal La327, derived from the spin configuration of (a). 
(c) La(2)$_{\rm b}$ sites around O$_{\rm vac}$s are dominated by the enhanced $H_{\rm int}$ (see arrows of different sizes). 
(d) Structure of La327 with Ni moments (red/blue arrows) corresponding to the area enclosed by the orange rectangle in (a). The light green La(1) and La(2) sites indicate the nearest neighbor sites of O$_{\rm vac}$. 
(e) Spatial distribution of La(2)$_{\rm a}$ (dark green) and La(2)$_{\rm b}$ (light green) around O$_{\rm vac}$s (gray) on the Ni-spinless channel above $T^*$, corresponding to (c) below $T^*$. 
The $H_{\rm int}$ at the La(2) is either parallel or antiparallel to the direction of $M_{\rm Ni}$ in this model, which is determined by the properties of the $A_{\rm hf}$ composed of transferred and/or dipolar fields. 
We temporarily draw the case dominated by the dipolar fields for simplicity.
}
\label{Ovac}
\end{figure*}

\newpage
\subsection{Influence of relative positional differences between the inner apical O$_{\rm vac}$ site and Ni-spin/spinless channels} 

Here we comment on the reason why we adopt the scenario that the inner apical O$_{\rm vac}$ site is located on the Ni-{\it spin} channel in Fig. \ref{SDW}. In order to obtain such a conclusion, we also consider the another possibility that inner apical O$_{\rm vac}$s are located on the Ni-{\it spinless} channels. It is possible that an O$_{\rm vac}$ can randomly come into any inner apical oxygen sites. Figure \ref{Ovac}(e) shows the case where all inner apical O$_{\rm vac}$s are on the Ni-{\it spinless} channels. In this case, $H_{\rm int}$ is canceled on the Ni-spinless channel even at the La(2)$_{\rm b}$ site. This is not consistent with the present NQR result, where there are few or no La(2)$_{\rm b}$ sites with $H_{\rm int}$ = 0. Therefore, most of inner apical O$_{\rm vac}$s are likely located on the Ni-spin channel. These results suggest that O$_{\rm vac}$s may also form a stripe pattern like the Ni-spin/spinless channels. Ni spins are induced on the channel with higher oxygen deficiency, corresponding to the Ni-spin channel. On the other hand, no Ni spins are induced in the channel with almost no oxygen deficiency, corresponding to the Ni-spinless channel. It is of interest in terms of how the possible spatial pattern of O$_{\rm vac}$s affects the SDW order at low pressure and the superconductivity at high pressure.

\end{document}